\documentclass[sigconf,nonacm]{acmart}

\microtypesetup{activate=false}
\setcopyright{none}
\AtBeginDocument{%
  }

\usepackage{tabularx}
\usepackage{xspace}
\usepackage{float}
\usepackage{placeins}
\usepackage{capt-of}
\usepackage{fancyvrb}
\usepackage{needspace}

\newcommand{\spara}[1]{\vspace{0.5em}\noindent\textbf{#1.}}
\newcommand{\system}{Kalypso\xspace}

\newcommand{\semsys}{SQPS\xspace}
\newcommand{\semsyss}{SQPSs\xspace}
\newcommand{\codedf}[1]{\textcolor{black}{#1}}
\newcommand{\codestr}[1]{\textcolor{blue!70!black}{#1}}

\newcommand{\semopfilter}{\textcolor{violet}{.sem\_filter}}
\newcommand{\semopjoin}{\textcolor{violet}{.sem\_join}}
\newcommand{\indexopsearch}{\textcolor{violet}{.index\_search}}
\newcommand{\semopmap}{\textcolor{violet}{.sem\_map}}
\newenvironment{workloadcode}{%
  \VerbatimEnvironment
  \Needspace{12\baselineskip}
  \begin{samepage}
  \begin{Verbatim}[
    fontsize=\scriptsize,
    commandchars=\\\[\],
    frame=single,
    framerule=0.4pt,
    framesep=1mm,
    numbers=left,
    numbersep=2pt,
    listparameters={\setlength{\topsep}{1.2em}\setlength{\partopsep}{0pt}}
  ]%
}{%
  \end{Verbatim}
  \end{samepage}
}
\newcounter{algorithm}
\renewcommand{\thealgorithm}{\arabic{algorithm}}
\newcounter{algoline}
\newlength{\algnumwidth}
\newlength{\algindent}
\newlength{\algbodywidth}
\newcommand{\algcalcwidth}[1]{%
  \setlength{\algbodywidth}{\columnwidth}%
  \addtolength{\algbodywidth}{-\algnumwidth}%
  \addtolength{\algbodywidth}{-\algindent}%
  \addtolength{\algbodywidth}{-#1em}%
}
\newenvironment{pseudoalgorithm}[1]{%
  \refstepcounter{algorithm}%
  \setcounter{algoline}{0}%
  \smallskip
  \noindent\rule{\columnwidth}{0.4pt}\par
  \noindent\textbf{Algorithm \thealgorithm} #1\par
  \noindent\rule{\columnwidth}{0.4pt}\par
}{%
  \noindent\rule{\columnwidth}{0.4pt}\par
  \smallskip
}

\newcommand{\alglineplain}[2]{%
  \noindent\makebox[\algnumwidth][r]{}\hspace{\algindent}%
  \algcalcwidth{#1}%
  \hspace*{#1em}\parbox[t]{\algbodywidth}{#2}\par
}

\newcommand{\algline}[2][0]{%
  \refstepcounter{algoline}%
  \noindent\makebox[\algnumwidth][r]{\scriptsize\thealgoline}\hspace{\algindent}%
  \algcalcwidth{#1}%
  \hspace*{#1em}\parbox[t]{\algbodywidth}{#2}\par
}

\begin{document}

\title{\system: Relational LLM Serving}

\author{Hojae Son}
\affiliation{%
  \institution{UMass Amherst, USA}
  \city{Amherst}
  \state{Massachusetts}
  \country{USA}
}
\email{hojaeson@umass.edu}

\author{Md Ashraful Islam}
\affiliation{%
  \institution{UMass Amherst, USA}
  \city{Amherst}
  \state{Massachusetts}
  \country{USA}
  }
\email{mdashrafulis@umass.edu}

\author{Huy Gia Cao}
\affiliation{%
  \institution{UMass Amherst, USA}
  \city{Amherst}
  \state{Massachusetts}
  \country{USA}
  }
\email{hcao@umass.edu}

\author{Hui Guan}
\affiliation{%
  \institution{UMass Amherst, USA}
  \city{Amherst}
  \state{Massachusetts}
  \country{USA}
}
\email{huiguan@cs.umass.edu}

\author{Marco Serafini}
\affiliation{%
  \institution{UMass Amherst, USA}
  \city{Amherst}
  \state{Massachusetts}
  \country{USA}
}
\email{marco@cs.umass.edu}

\renewcommand{\shortauthors}{Son et al.}

\begin{abstract}
Large language models are increasingly used as semantic operators for filtering, extracting, ranking, joining, and transforming unstructured data. Existing semantic query processing systems invoke request-centric LLM serving systems that are unaware of the query plan, leaving substantial performance opportunities unused. This paper introduces relational LLM serving, an abstraction that makes LLM serving aware of semantic query structure while preserving query semantics and output accuracy. The key opportunity is pipelined execution across semantic operators: when intermediate tuples flow directly from one operator to the next, their KV-cache state can be reused instead of recomputed.

We present \system, a relational LLM serving system that exposes an API for semantic query plans and executes them using an adaptive, memory-aware scheduling algorithm. 
\system addresses a new online scheduling problem in which pipelined operator execution is coupled with GPU memory pressure management to reuse KV-cache state in the serving engine before eviction.
Its scheduler continuously adjusts memory allocations to balance upstream parallelism, downstream progress, and GPU utilization. 
Our evaluation shows that \system improves query completion time over baselines using request-centric LLM serving, with speedups up to 4.57$\times$ across diverse workloads, demonstrating that query-aware LLM serving can substantially improve the efficiency of semantic query execution.

\end{abstract}

\makeatletter
\def\@mkabstract{\bgroup
  \ifx\@abstract\@lempty\else
  {\phantomsection\addcontentsline{toc}{section}{Abstract}%
    \if@ACM@journal
       \everypar{\setbox\z@\lastbox\everypar{}}\small
    \else
      \section*{\abstractname}%
    \fi
   \ignorespaces\@abstract\par}%
  \fi
  \par\kern\medskipamount
  \egroup}
\maketitle
\makeatother

\vspace{-8pt}
\section{Introduction}
\label{sec:intro}

Large language models (LLMs) have emerged as a foundational abstraction for processing unstructured data in modern data management systems. Operators such as semantic filtering, extraction, ranking, and transformation can now be expressed declaratively over natural language inputs, allowing unstructured corpora to be queried and manipulated using relational-style operators \cite{patel2025semantic,liu-etal-2024-suql}. This shift enables a new class of data-intensive applications in analytics, retrieval-augmented reasoning, and agentic AI.
To integrate LLM inference into query processing, recent \emph{semantic query processing systems} (\semsyss) introduce \emph{semantic operators}, which extend traditional relational algebra operators to encapsulate LLM-inference with user-defined prompts~\cite{patel2025semantic,liu2025palimpzest,jo2024thalamusdb, shankar2024docetl}.
These operators act on tables whose rows may contain unstructured text fields---for example, product descriptions, clinical notes, or contract clauses---and each row's content is serialized into an LLM prompt for processing.
For example, a semantic filter can retain products relevant to the natural-language query \emph{"\{product.review\} Does this review criticize battery life?"}~\cite{ovcharenko2026sempipes}, while a semantic join can match patients to clinical trials using a natural-language predicate across tables~\cite{jin2024matching}.
\emph{Semantic queries} combine multiple semantic operators into a query plan.

The high cost of LLM inference represents the main bottleneck in the execution of semantic queries.
\semsyss invoke inference requests on LLM models using LLM serving systems~\cite{kwon2023efficient,zheng2024sglang, yu2022orca, agrawal2023sarathiefficientllminference, aminabadi2022deepspeed}.
These systems optimize request scheduling, GPU memory management, and prefix caching through optimizations such as paged attention, prefix sharing, or continuous batching.
However, they are request-centric: they see the workload as a sequence of separate inference requests and lack a high-level understanding of the semantic query.

On the other hand, existing SQPS are aware of semantic queries but use this knowledge to \emph{reduce} the number of LLM inference requests or \emph{introduce LLM model approximations} rather than \emph{making LLM serving itself more efficient}, which is the focus of this paper.
Much prior work on semantic query execution has focused on reducing the number of expensive oracle LLM calls through cascaded execution with cheaper proxy models~\cite{patel2025semantic,shankar2026task,Chung_2026,ong2025routellmlearningroutellms}, similarity-based pruning and pipeline decomposition~\cite{jo2024thalamusdb,shankar2024docetl,Hu_2025,ram2023context}, request reordering to improve prefix sharing~\cite{liu2024relationalworkloads}, and query optimization to find efficient plans~\cite{liu2025palimpzest, russo2025deep}.
These techniques are effective but they still leave running inference on the LLM  (or oracle LLM for implementations using cascading) as the major performance bottleneck. 
These optimizations are complementary to increasing LLM serving efficiency, and introduce a tradeoff between the accuracy of the query results and query running time, which is an orthogonal concern.

In this paper, we propose \emph{relational LLM serving}, an approach that optimizes LLM serving on tabular data by making it aware of the semantic query plan.
A relational serving layer sits between \semsyss and the underlying request-centric inference engine: 
rather than issuing LLM requests independently, the SQPS delegates a query plan, and the serving layer decides which tuples enter which operators, when requests launch, and how KV-cache memory is allocated. This scheduling is an execution optimization that does not change query semantics or output accuracy.

The key benefit of relational LLM serving is \emph{improving the KV cache hit rate through pipelining}.
Semantic operators process tuples by including them in the prompt of the associated LLM requests.
The prefill cost of computing KV cache entries for tuples is a major bottleneck in LLM serving for these workloads.
By knowing the query plan, relational LLM serving systems can leverage pipelining opportunities, where an operator is invoked on intermediate tuples as soon as they are produced.
When a tuple is processed by an operator and then immediately passed to the next operator, the corresponding KV cache prefix can be reused and it does not need to be recomputed.
This is in contrast with the operator-at-a-time strategy common in existing \semsyss, which materialize intermediate results without using  pipelining.

This paper describes the design and implementation of \system, the first relational LLM serving system.
We start by defining an \emph{API for relational LLM serving}.
Recent \semsyss have proposed many efficient semantic operator semantics and implementations.
\system exposes a simple API to define semantic operator implementations and query plans that is general enough to support these optimizations, for example enabling the use of external proxy models or tools such as vector indexes, but also expressive enough to expose pipelining opportunities.

Leveraging pipelining for semantic queries requires addressing a challenging and novel problem: \emph{online scheduling of dependent (pipelined) operators under bounded KV-cache capacity}. 
To achieve high GPU utilization, a scheduler must process many tuples concurrently, which induces GPU memory pressure.
If the scheduler runs too many upstream operator instances, memory pressure can evict cached prefixes before all its dependent downstream operators can reuse them.
Conversely, if upstream operators run with too little parallelism, downstream operators may be starved for input, reducing GPU utilization.
The right degree of parallelism is also influenced by data- and query-dependent factors, such as filter selectivity and join fanout, which are revealed only \emph{online}, during query execution.
 
\system addresses this problem through \emph{adaptive memory-aware scheduling}: it reserves a memory budget to each operator it launches, which is used to cache its prefix.
It then uses memory utilization information to control how many operator instances can run concurrently without undesired prefix evictions.

Making \emph{memory management robust} is also challenging.
The memory required by a semantic operator is not known statically, because LLM requests may generate a variable number of tokens. 
The memory allocation algorithm of \system needs estimates of each operator's memory demand for a tuple before executing it. Overestimating this demand reduces parallelism by reserving memory unnecessarily, whereas underestimating it may force an operator to be rerun with a larger allocation, and can cause unplanned evictions of prefixes.
\system also offers multiple options for pinning KV cache memory and introduces mechanisms for deadlock detection and recovery.

We evaluate \system on four semantic-query workloads spanning fact
verification, biomedical entity matching, medical error correction, and
contract entailment. Compared with request-centric execution in existing
\semsyss, \system reduces end-to-end query completion time by up to 4.57$\times$
while issuing a similar number of LLM calls. Controlled ablations isolate
the effects of \system's key design choices.

In summary, this paper makes the following contributions:

\begin{itemize}
    \item We propose a new architecture (Section~\ref{sec:overview}) and a general API for relational LLM serving that supports existing semantic operator implementations (Section~\ref{sec:API}). 
    \item We describe an adaptive scheduler for relational LLM serving, which supports pipelining and high parallelism while controlling cache evictions (Section~\ref{sec:scheduler}).
    \item We discuss memory management techniques such as memory estimation, pinning, and deadlock detection and recovery (Sections~\ref{sec:robustness}).
    \item We compare the performance of \system to existing SQPS, which rely on materialization and request-centric LLM serving systems (Section~\ref{sec:evaluation}).
\end{itemize}

\section{Background and Motivation}

\subsection{Background}
Autoregressive LLM inference~\cite{vaswani2017attention} consists of two phases.
During \emph{prefill}, the model processes all prompt tokens in parallel and materializes per-layer key-value (KV) tensors.
During \emph{decode}, the model generates one token at a time, attending over all previously computed KV tensors.
To avoid recomputing these tensors at every step, inference engines store them in GPU memory as the \emph{KV cache}.
The KV cache grows linearly with sequence length and is often the primary memory bottleneck in LLM serving.

\spara{Prefix caching}
When multiple requests share an identical prompt prefix, the corresponding KV tensors are identical and can be computed once and reused, avoiding redundant prefill.
Modern serving systems such as vLLM~\cite{kwon2023efficient} and SGLang~\cite{zheng2024sglang} support this through automatic prefix caching: if a new request's prompt begins with the same tokens as a recently served request, the engine reuses the cached KV entries and only computes the novel suffix.
However, this reuse is \emph{opportunistic}: once a request completes, its KV-cache blocks become reclaimable and may be evicted at any time under memory pressure. There is no mechanism to guarantee that a prefix remains resident for a future request.

\spara{vLLM memory model}
\system builds on vLLM~\cite{kwon2023efficient}, which manages KV-cache memory as a global pool of fixed-size blocks, analogous to virtual memory pages.
Blocks are allocated from the pool when a request is admitted and returned when it completes; if insufficient blocks are available, new requests are deferred.

\spara{Prefix sharing in semantic operators}
Semantic operators naturally exhibit prefix sharing.
When consecutive operators in a query plan are invoked on the same tuple, their prompts often begin with the same system instructions and tuple context.
Only the trailing task instruction changes from one operator to the next.
The KV-cache entries for this shared prompt prefix are therefore valid for the downstream operator, as long as they remain resident in GPU memory.

\begin{figure}[t]
    \centering
    \includegraphics[width=1\linewidth]{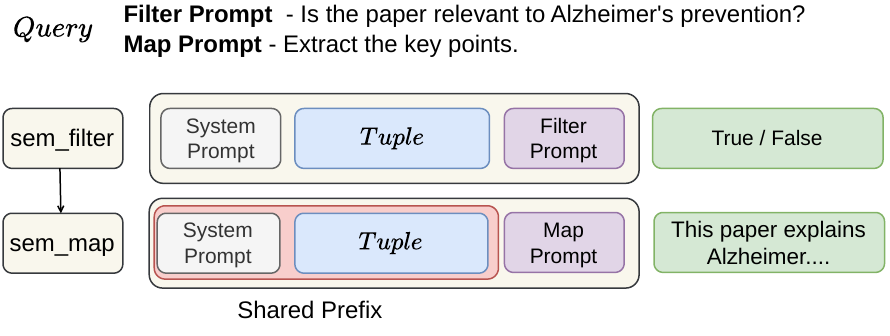}
    \caption{Prompt structure for a filter $\rightarrow$ map query.}
    \Description{Diagram showing a filter followed by a map on the same context. The filter and map prompts share the system prompt and context prefix, while the operator-specific instruction changes.}
    \label{fig:prompt-structure}
\end{figure}

Figure~\ref{fig:prompt-structure} illustrates this structure for a
filter $\rightarrow$ map query over a paper dataset. Let $S$ denote the
system prompt and $C(t)$ the paper text for tuple $t$. The filter prompt
is $[S \mid C(t) \mid I_{\mathrm{filter}}]$, where
$I_{\mathrm{filter}}$ asks whether the paper is relevant to Alzheimer's
prevention. The downstream map prompt is
$[S \mid C(t) \mid I_{\mathrm{map}}]$, where $I_{\mathrm{map}}$ asks to
extract the key points. The shared reusable prefix is therefore
$[S \mid C(t)]$. The filter's boolean output only determines whether the
tuple reaches \texttt{sem\_map}; it is not included in the map prompt.
If the shared prefix remains cached, the map reuses its KV-cache state
and only prefills the map-specific suffix.

\subsection{Motivation}

\spara{The importance of KV cache hits}
Existing \semsyss execute queries operator-at-a-time, materializing intermediate results between operators.
This design leads to KV-cache evictions under memory pressure, forcing downstream operators to recompute prefixes.
We now quantify the impact of these cache misses on end-to-end query execution time.

We run a semantic filter on Lotus using Llama-3.2-3B on an NVIDIA A16 GPU (16~GB).
Each tuple is padded to 750 tokens, consuming approximately 84~MB of KV-cache memory per tuple (112~KB/token $\times$ 750 tokens).
After model weights and runtime overheads, roughly 6~GB remains for KV cache, enough to hold at most ${\sim}$70 tuples simultaneously.

\begin{figure}[!t]
\centering
\captionsetup{aboveskip=2pt,belowskip=-10pt}
\includegraphics[width=1\linewidth]{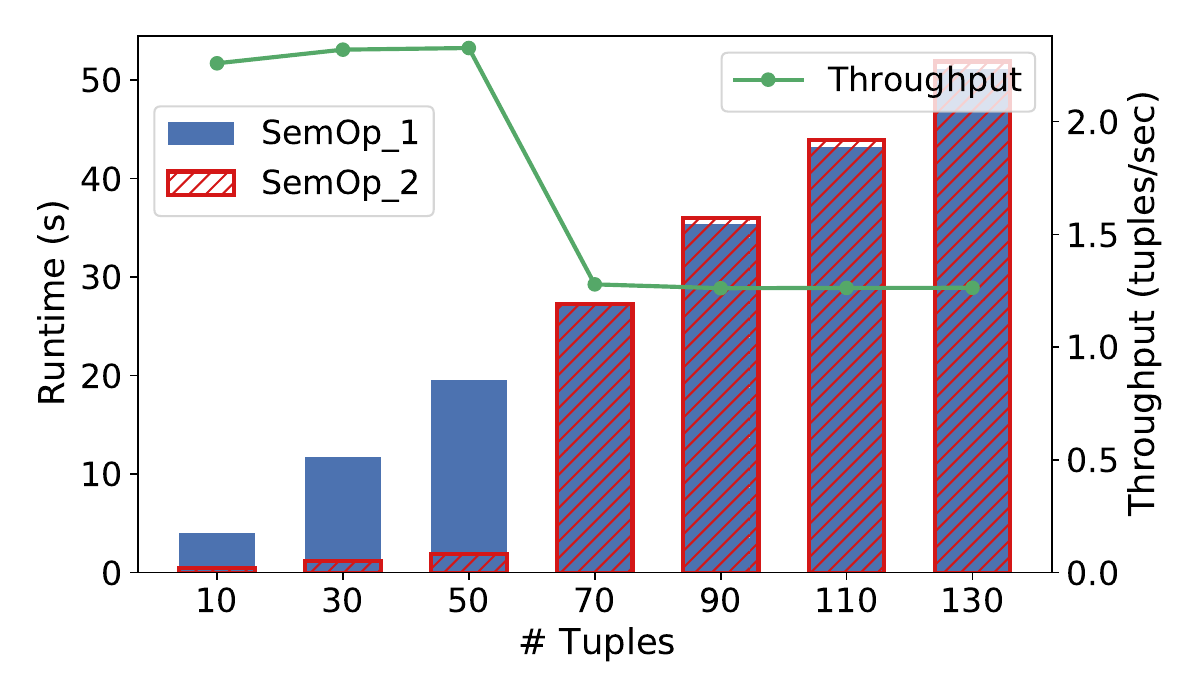} 
\caption{When the input table exceeds available GPU KV-cache memory, downstream operators incur cache misses and must recompute the prefix.}
\Description{Line plot showing downstream operator runtime increasing sharply after the working set exceeds available KV-cache memory.}
\label{fig:cache_miss_exp}
\end{figure}

With this setup, we run a query consisting of two concatenated filters on the same table, varying the number of tuples in the table.
We adopt the approach of Lotus, which executes the first operator on the entire table and fully materializes the intermediate table before executing the second operator.
The results are reported in Figure~\ref{fig:cache_miss_exp}.
The first operator prefills the KV cache for each input tuple, so its running time scales linearly as the number of tuples increases.
When the table has fewer than 70 tuples, the second operator does not need to execute the prefill again because it hits the KV cache, so its running time is minimal.
However, if the table is larger than 70 tuples, the LLM serving system starts evicting tuples before the first operator completes, inducing cache misses for the second operator.
The second operator now needs to execute prefill on the tuple again and takes the same time as the first one, which results in a spike in end-to-end running time and a drop in throughput. 
Using an MRU eviction policy instead of the default LRU would improve cache hits only by a constant factor, whose impact in the end-to-end query runtime decreases as the size of the table increases.

\spara{The challenge of memory-aware pipelining}
The experiment above shows that an operator-at-a-time execution strategy cannot achieve prefix reuse under memory pressure.
Pipelining operators---scheduling the downstream operator on a tuple promptly after the upstream produces it --- can potentially exploit the serving engine's opportunistic prefix caching, but it needs to strike a balance between multiple objectives.
High GPU utilization requires processing many tuples concurrently, which introduces high memory pressure.
In this regime, the serving engine must evict cached blocks to admit new requests.
Reliable cross-operator reuse therefore requires a \emph{memory-aware scheduler}, which controls operator launches  based on the available memory to consume reusable prefixes before memory pressure evicts them.
Admitting too many upstream tasks increases memory pressure and can evict reusable prefixes, while admitting too few upstream tasks can generate too little work for downstream progress. This is the challenging scheduling problem we address in this paper.

\begin{figure}[!t]
	\centering
	\includegraphics[width=\linewidth]{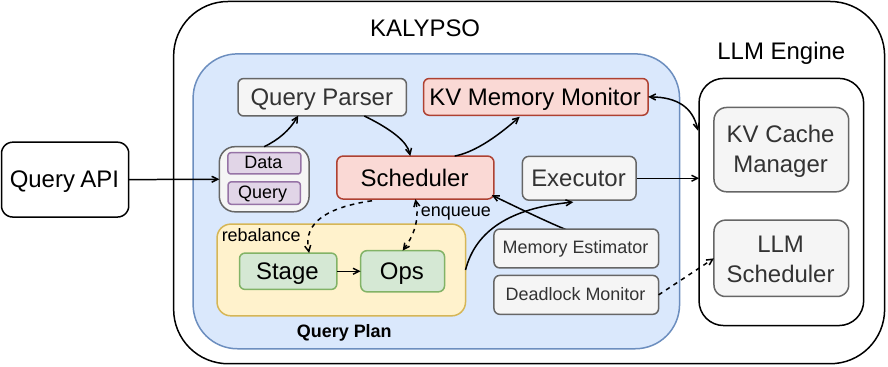}
	\caption{\system architecture. \system sits between the query client and the LLM engine. Solid arrows indicate data flow; dashed arrows indicate scheduling control.}
	\Description{Relational LLM Serving System Architecture.}
	\label{fig:system_overview}
\end{figure}

\begin{table*}[t]
\centering
\small
\setlength{\tabcolsep}{3pt}
\renewcommand{\arraystretch}{0.92}
\caption{Characteristics of common semantic operators and their execution properties.
$M(\cdot,l)$ denotes LLM evaluation under natural-language specification
$l$, and $G(\cdot,l)$ denotes label generation for classification.}
\label{tab:semantic-ops}
\begin{tabularx}{\textwidth}{l >{\raggedright\arraybackslash}p{0.34\textwidth} >{\raggedright\arraybackslash}p{0.19\textwidth} >{\centering\arraybackslash}p{0.13\textwidth} >{\centering\arraybackslash}p{0.06\textwidth} >{\centering\arraybackslash}p{0.06\textwidth}}
\toprule
Operator & Description & Logical Definition & Pipelining & Predicate & CP \\
\midrule

\texttt{sem\_filter}(l)
& Select tuples satisfying a language predicate
& $\{t \in T \mid M(t,l)=1\}$
& Y
& Y
& N \\

\texttt{sem\_map}(l)
& Transform each tuple via a language instruction
& $\{M(t,l) \mid t \in T\}$
& Y
& N
& N \\

\texttt{sem\_join}(l)
& Join two relations via language predicate
& $\{(t_i,t_j)\mid M((t_i,t_j),l)=1\}$
& Y
& Y
& Y \\

\texttt{sem\_classify}(l)
& Assign category labels to each tuple
& $\{(G(t,l)) \mid t \in T\}$
& Y
& N
& N \\

\texttt{sem\_agg}(l)
& Aggregate multiple tuples using a language reducer
& $M(\{t_1,\dots,t_n\},l)$
& N
& N
& N \\

\texttt{sem\_topk}(l,k)
& Select top-$k$ tuples according to language ranking
& $\operatorname{Top}_k(T,l)$
& N
& N
& N \\

\bottomrule
\end{tabularx}
\end{table*}

\section{Overview of \system}
\label{sec:overview}

We now give an overview of \system, a relational LLM serving system that executes semantic queries and maximizes cross-operator KV-cache reuse.
The main components are shown in Figure~\ref{fig:system_overview}.

\system acts as a layer between the query client, which could be a \semsys like Lotus, and an LLM engine, such as vLLM.
The query client specifies a query plan consisting of semantic operators.
These are implemented by the \semsys as user-defined functions (UDFs) through \system's API.
Operators invoke LLM requests through \system's Executor, which acts as a wrapper to keep track of request completions and context (see Section~\ref{sec:API}).

\system's scheduler coordinates pipelined query execution by launching operators, which are organized into pipelined fragments called stages.
The scheduler is memory-aware: it uses GPU memory occupancy information from the underlying LLM engine to decide which operators to launch concurrently, control memory pressure, and ensure that cached prefixes are not evicted while they are still being consumed.
\system's scheduler acts as an admission control system for operators, while
the actual execution of LLM requests is scheduled on the GPU by the LLM engine (see Section~\ref{sec:scheduler}).

The system includes additional components to manage memory.
The memory estimator is used to predict the number of tokens required to run an operator on a tuple.
Optionally, this information can be used to pin GPU memory in the LLM engine's KV cache manager.
The deadlock manager ensures progress when pinning introduces deadlocks (see Section~\ref{sec:robustness}).

This architecture allows \system to execute semantic queries as memory-aware pipelines: the scheduler continuously launches new operators as memory becomes available, supporting pipelining and keeping GPU utilization high throughout query execution while avoiding undesired cache evictions.

\section{\system API}
\label{sec:API}
The \system API can be used by \semsyss such as Lotus to register their semantic operators and their implementations with \system.
Query clients can then submit query plans using these operators to \system, which takes care of their execution.

\spara{Implementing semantic operators}
With \system, \semsyss implement each operator as a User-Defined Function (UDF).
The UDF logic is opaque to \system, except that the LLM requests it serves must be invoked through a wrapper.

\system requires each operator implementation to declare an execution contract specifying three properties, which are essential for scheduling query execution: \emph{(1- Pipelining)} can the operator be pipelined or it is blocking? \emph{(2- Predicate)} can the operator prune its input tuple, interrupting downstream processing for that tuple? \emph{(3- Cartesian Product)} does the operator join tuples from two tables? 
Table~\ref{tab:semantic-ops} classifies semantic operators introduced in~\cite{patel2025semantic} according to \system's API.

A \emph{pipelining} operator takes one tuple as input and produces at most one tuple as output.
For example, semantic filter and map operators are pipelining.
Operators that are not pipelining are treated as blocking:
\system materializes their entire input table before executing them.
Aggregation and top-$k$ are examples of blocking operators because they combine data from multiple tuples.

A \emph{predicate} operator returns a boolean value indicating whether its output should be forwarded to downstream operators.
If the predicate prunes a tuple, the memory caching its prefix can be freed.
Operators that are not predicates always forward their output to the dowstream operator.
Semantic filter operators are a common example of predicate operators.

A \emph{Cartesian Product (CP)} operator combines tuples from a left and a right table. 
The execution of the actual Cartesian product must be performed by calling a \system procedure.
This gives the system control over the scheduling of downstream operators, as we will describe.
The left-side table must be either a static input table or the output of another operator, since \system supports pipelining on the left side of Cartesian products.
The right-side table is another static input table.
This allows implementing joins as a CP operator followed by a filter operator.

\spara{Prompt format and prefix sharing}
\system enables prefix sharing by scheduling dependent pipelined operators before the relevant cached prefixes are evicted.
It is up to the \semsys's operator implementations to format tuples and, more generally, the prompts to the LLM in a way that maximizes prefix sharing opportunities among pipelined operators, as shown in the example of Figure~\ref{fig:prompt-structure}.

\spara{Cascading, vector indexes, and other optimizations}
Defining operators as UDFs allows a wide range of efficient operator implementations that are common in \semsyss.
For example, filter operators may be implemented as a cascade of a cheaper proxy LLM and a more expensive oracle LLM, which is invoked only if the proxy has low confidence in its decision.
Operator UDFs can also use external tools besides LLMs, such as embedding models or vector indexes.

The API also supports optimized join implementations that avoid full Cartesian products, which are expensive.
CP operators can optionally include a UDF that takes each left-side tuple and returns a smaller subset of right-side tuples to be joined. 
The UDF can be used, for example, to integrate a vector index as a cheaper proxy implementation of a join, an implementation that we call Indexed Cartesian Product (ICP).
Suppose that the tuples in the right-side table of the join are indexed using a vector index or vector database.
The UDF can use each left-side tuple as a query to the vector index.
Then, the UDF can use the response to select the right-side tuples that should be joined with the left-side tuple.
The final Cartesian product between the tuples is executed by \system.

\enlargethispage{2\baselineskip}
\spara{Queries}
At query execution time, \system receives static \emph{left-deep} query plans.
It schedules the execution of the operators in the plan to maximize GPU utilization, KV-cache hits, and prefix sharing.
It then returns the output of the query to the client.

\section{Scheduling}
\label{sec:scheduler}

We now describe how \system schedules query execution and LLM inference calls to achieve pipelining and KV-cache reuse.

\begin{figure*}[t]
	\centering
	\includegraphics[width=\linewidth]{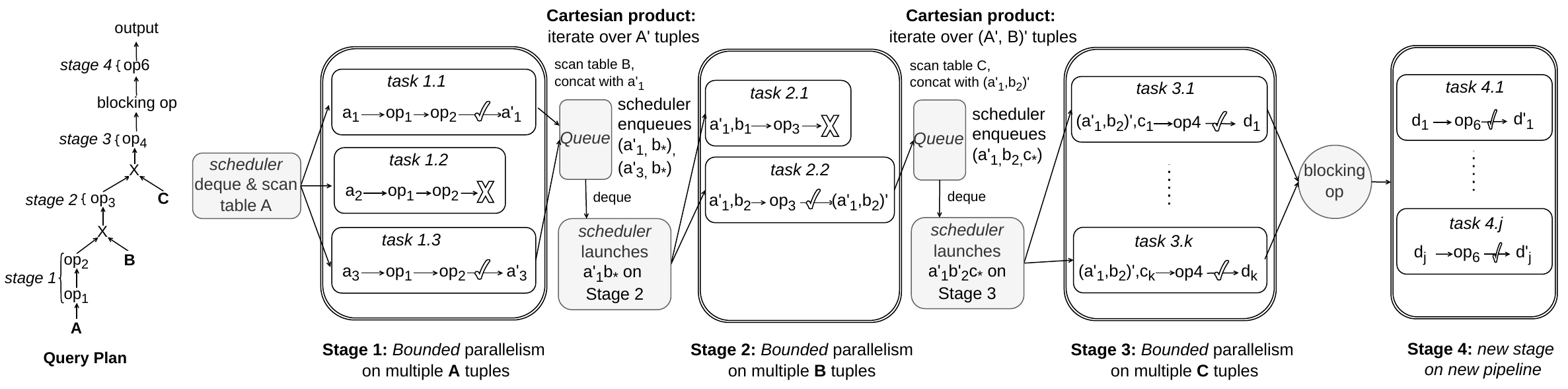}
	\caption{Executing a semantic query plan with pipelining.}
	\Description{Execution timeline illustrating pipelined stages, dependent tasks, and cache-aware scheduling under memory pressure across a semantic query plan.}
	\label{fig:scheduler_execution}
\end{figure*}

\subsection{Executing Query Plans}
The query plan is processed by the query parser, which groups operators into \emph{pipelines} (separated by blocking operators) and further into \emph{stages} within each pipeline (separated by Cartesian products such as joins).
Each stage is a sequence of one or more operators that execute sequentially on a tuple---for example, a filter followed by a map.
Running a stage on one tuple constitutes a \emph{task}, the atomic scheduling unit.

Consider the left-deep query plan on the left of Figure~\ref{fig:scheduler_execution} as a running example.
The plan has four stages: the first three constitute one pipeline, the fourth in is a separate pipeline.
Tuples flow within tasks and across stages of the same pipeline without materialization: downstream tasks are scheduled to reuse the upstream task's prefix through the serving engine's prefix cache.

The scheduler runs multiple tasks concurrently---each task executes a stage on one input tuple.
In the first stage, the scheduler scans table $A$ and launches three tasks.
Task~1.1 takes tuple $a_1$ and produces $a'_1$ (extending the tuple with generated content); task~1.2 filters out $a_2$; task~1.3 similarly takes $a_3$ and produces $a'_3$.

When a task completes, its output is either materialized, if the task belongs to the final stage of a pipeline, or pushed into a queue for the next stage's Cartesian product.
The Cartesian product combines output tuples with a right-side table to create \emph{dependent tasks}.
In Figure~\ref{fig:scheduler_execution}, $a'_1$ enters the queue for the second stage; the Cartesian product combines it with tuples from table $B$, producing tasks~2.1 and~2.2 as dependent tasks of task~1.1.
These dependent tasks are launched by the scheduler.
Figure 4 says that the queue contains the outputs of the tasks. The rest of the paper, including Figure 5 and 6 and the scheduling algorithm, say that the queue contains the new tasks.

\subsection{Memory-Aware Scheduling}
\label{subsec:mem-based-schedule}
The query execution framework described previously and illustrated in the example of Figure~\ref{fig:scheduler_execution} still needs a scheduling algorithm to decide which task should be launched and when.
To enable prefix reuse, the scheduler should leverage pipelining, that is, prioritize launching dependent downstream tasks can reuse the KV-cache state of a completed task without recomputation.
For example, in Figure~\ref{fig:scheduler_execution}, after completing task~1.1, the scheduler should prioritize launching its dependent tasks.
However, simply prioritizing dependent tasks is not sufficient to achieve high cache hit rates: ideally, all dependent tasks should be launched \emph{before memory pressure evicts the prefix} for $a'_1$.
If the scheduler launches too many concurrent tasks, it could create memory pressure that could lead to the eviction of the prefix for $a'_1$ before the dependent tasks execute.
Therefore, besides doing basic pipelining, the scheduler also needs to control the launching of tasks, and their associated memory pressure, in order to control the timing of cache evictions.

\system's scheduler \emph{times the launching of tasks based on memory availability}, so that cached prefixes that are still needed by dependent tasks don't need to be evicted.
It controls memory availability by keeping track of the size of the cached prefixes that are still needed.
It also bounds the amount of memory each task is allowed to use, using a memory estimator, and optionally offers support for memory pinning, as we discuss in Section~\ref{sec:robustness}.

\subsection{The \system Scheduling Algorithm}
\label{sec:scheduling-problem}
Pipelined execution creates a unique online scheduling problem because KV-cache reuse couples operator dependencies with GPU memory pressure. 
This makes it challenging to achieve high parallelism.
When a tuple crosses an operator boundary, \system must reserve memory and admit downstream operators while avoiding that memory pressure evicts the reusable prefix. 
Admitting too many early-stage tasks can exhaust memory and limit parallelism for downstream stages, while admitting too few can leave downstream stages without enough input to keep the GPU busy.

We now illustrate this problem and introduce the insights that motivate the \system scheduling algorithm by first discussing three baseline algorithms: a simple sequential algorithm, which creates minimal memory pressure but underutilizes the GPU, a parallel depth-first algorithm, which can \emph{starve} upstream operators and reduce pipeline throughput, and a parallel breadth-first algorithm, which can \emph{saturate} KV-cache capacity by admitting too much upstream work. 
Each algorithm addresses some aspects of the scheduling problem but also reveals some of its challenges.
We then combine these insights and ideas in the \system scheduling algorithm.

\spara{Sequential depth-first execution}
The simplest way to execute queries would be for the scheduler to launch tasks sequentially in a nested loop, completing each task and its dependents before moving on to maximize prefix reuse.
This, however, would run at most one LLM request at any point in time, which is not sufficient to fully utilize GPU resources.
In Figure~\ref{fig:scheduler_execution}, this policy runs task~1.1, then tasks~2.1 and~2.2, and then all task~3.x instances dependent on task~2.2, one for each tuple in table $C$.
When the next operator is blocking, such as \texttt{SemTopK} or \texttt{SemAgg}, the pipeline stalls until that operator has consumed its dependent inputs and produced output.
After this barrier, task~4.x starts a new single-stage pipeline using the available memory capacity.
Although simple, this policy keeps at most one task active and underutilizes the GPU.
Therefore, the \system scheduler launches multiple tasks in parallel, but keeping high GPU utilization has some challenges, as we now discuss.

\spara{Parallel depth-first execution and starvation}
A natural extension of the sequential depth-first policy is to parallelize the last stage of a pipeline by assigning it a large static memory budget and launching as many tasks as possible within that budget.
This strategy tries to finish dependent tasks as soon as possible: once an upstream task has produced downstream tasks, it launches many parallel last stage tasks so that the upstream tuple's reusable prefix can be consumed quickly and then evicted.

\begin{figure}[t]
    \centering
    \includegraphics[width=1 \linewidth]{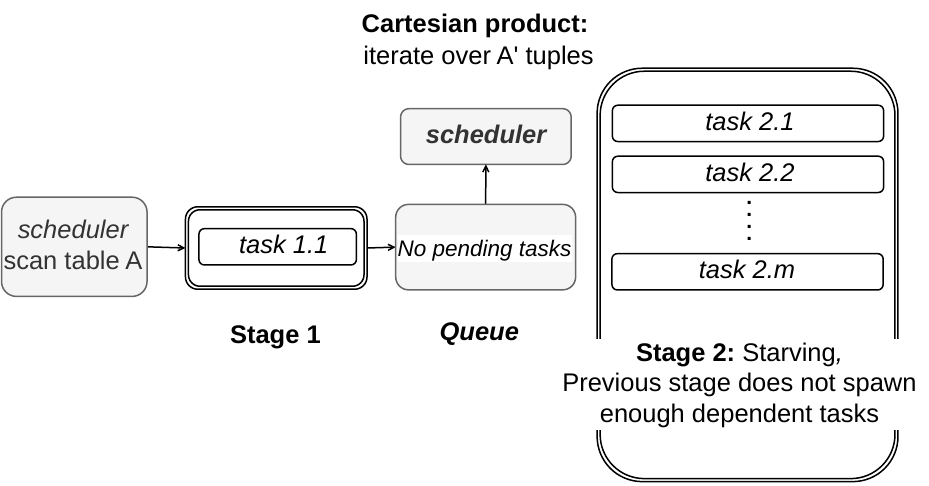}
    \caption{A parallel depth-first policy can starve downstream stages.}
    \Description{Timeline showing a parallel depth-first scheduling policy that repeatedly prioritizes later-stage tasks, causing earlier-stage tasks to wait and reducing the supply of future downstream work.}
    \label{fig:starvation}
\end{figure}

The limitation of this policy is that it can \textbf{starve} the
later stages. Figure~\ref{fig:starvation}
illustrates this effect in a two-stage pipeline.
In the figure, task~1.1 has just completed and all dependent tasks of tasks~1.1 are immediately launched in parallel, leaving the queue empty.
When these tasks complete, the earlier stage runs task~1.2 in isolation, since only the last stage is parallelized, which underutilizes GPU resources.
Meanwhile, the last stage is starving because it has no work to do.
This policy minimizes the amount of memory dedicated to caching prefixes, freeing up the rest of the memory for running a large number of last-stage tasks.
However, it can underfeed the pipeline since upstream stages are not given enough memory budget run tasks in parallel and produce downstream work.

\spara{Parallel breadth-first execution and saturation}
A solution to the problem of starvation is to statically allocate more memory to the first stage so that it can run multiple parallel tasks.
This breadth-first strategy admits more upstream tasks
concurrently, exposing parallelism and creating a larger pool of dependent
downstream work and avoiding starvation.
However, if these static memory budgets are not tuned carefully, this policy can \textbf{saturate} the KV-cache budget with admitted upstream work. 
Figure~\ref{fig:saturation} illustrates this effect: many
early-stage tasks are admitted because the stage has a large budget.
When they complete, the queue fills up and the reusable prefixes stay cached until all their dependent tasks are completed.
This constraints the amount of memory available for launching later-stage tasks. 
Thus, breadth-first parallelism increases the supply of work, but it can overcommit memory to early stages and block progress at later stages.

\spara{The need for adaptive scheduling}
The starvation and saturation examples show that no fixed allocation of
stage memory budgets is robust across semantic query workloads. The right
budget split depends on properties such as operator selectivity and join fanout, i.e., how many right-side tuples join with each left-side tuple.
With high selectivity, few tuples survive, or with low fanout, each tuple
creates little downstream work, so the later stages starve for ready tasks;
with low selectivity and high fanout, many tuples survive and create large
amounts of downstream work, so admitted upstream work can saturate the KV-cache budget.
\system therefore adapts stage budgets online, shifting budget toward
earlier stages when the pipeline is starved for new downstream work and
toward later stages when admitted upstream work is saturating the KV-cache budget. 
We now detail how \system detects these conditions and reassigns budget across stages during execution.

\spara{The \system scheduling algorithm} \label{sec:scheduling-algo}
The previous algorithms show why the scheduling algorithm of \system has three goals: (1) \emph{avoid starvation}, where later
stages run out of ready dependent tasks because upstream stages do not receive
enough budget, (2) \emph{avoid saturation}, where upstream stages fill up the KV-cache and leave too little budget for downstream tasks to make progress, and (3) \emph{adapting online} to the characteristics of different queries and datasets.
To achieve these goals, the scheduler assigns a dynamic memory budget to each stage, tracking per-stage queue
pressure to determine when new tasks can be launched, and adapts these budgets online by rebalancing across operators.

\begin{figure}[t]
\footnotesize
\begin{pseudoalgorithm}{\system's Scheduling Algorithm (Single Pipeline).}
\label{alg:scheduler_pseudocode}

\algline{\textbf{procedure} \textsc{Schedule}$(\mathit{stages})$} 
\algline[1]{$\mathit{running} \gets \emptyset$, $\mathit{out} \gets [\ ]$}
\algline[1]{$\mathit{firstStage} \gets \mathit{stages}[0]$}
\algline[1]{$\mathit{tasks} \gets \mathit{firstStage}.\mathit{inputTable()}$}
\algline[1]{$\mathit{firstStage}.\mathit{waiting.enqueue}(\mathit{tasks})$}
\algline[1]{\textbf{while} $\mathit{stages.hasWaitingTasks()}$ \textbf{or} $\mathit{running}\neq\emptyset$ \textbf{do}}
\algline[2]{\textsc{Launch}$(\mathit{stages},\mathit{running})$}
\algline[2]{\textbf{for each} $(\mathit{stage}, \mathit{task}, \mathit{result}) \in \mathit{running.completedTasks()}$ \textbf{do}}
\algline[3]{$\mathit{running.delete(task)}$}
\algline[3]{\textsc{Complete}$(\mathit{stage}, \mathit{task}, \mathit{result})$}
\algline[2]{\textsc{Rebalance}$(\mathit{stages})$}
\algline[1]{\textbf{return} $\mathit{out}$}
\alglineplain{0}{}

\algline{\textbf{procedure} \textsc{Launch}$(\mathit{stages},\mathit{running})$}
\algline[1]{\textbf{for each} $\mathit{stage} \in \mathit{stages}$ \textbf{do}}
\algline[2]{\textbf{while} $\mathit{stage}.\mathit{waiting}\neq\emptyset$ \textbf{do}}
\algline[3]{$\mathit{task} \gets \mathit{stage}.\mathit{waiting.peek()}$}
\algline[3]{\textbf{if} $\mathit{memoryManager}.\mathit{admit}(\mathit{stage, task})$ \textbf{then}}
\algline[4]{$\mathit{task} \gets \mathit{stage}.\mathit{waiting.dequeue()}$}
\algline[4]{$\mathit{memoryManager.allocateMemory(stage,task)}$}
\algline[4]{$\mathit{running}.\mathit{add}(\mathit{task})$}
\algline[4]{$\mathit{launchWorker(task)}$}
\algline[3]{\textbf{else break}}
\alglineplain{0}{}

\algline{\textbf{procedure} \textsc{Complete}$(\mathit{stage}, \mathit{task}, \mathit{result})$}
\algline[1]{\textbf{if} $\mathit{result.retry()}$ \textbf{then}} \label{ln:retry-start}
\algline[2]{$\mathit{task.updateBudget}(\mathit{result})$}
\algline[2]{$\mathit{stage.waiting.enqueue(task)}$}
\algline[2]{\textbf{return}} \label{ln:retry-end}
\algline[1]{\textbf{if} $\mathit{result.filtered()} = \mathit{false}$ \textbf{then}}
\algline[2]{\textbf{if}  $\mathit{stage} \neq \mathit{stages.lastStage}()$ \label{ln:cp-start} \textbf{then}}
\algline[3]{$\mathit{nextStage} \gets \mathit{stages}[\mathit{stage}.\mathit{id}+1]$}
\algline[3]{$\mathit{childrenTasks} \gets \mathit{ExecuteCP}(\mathit{result},\mathit{nextStage.inputTable()})$}
\algline[3]{$\mathit{nextStage}.\mathit{waiting.enqueue}(\mathit{childrenTasks})$}
\algline[3]{$\mathit{memoryManager}.\mathit{trackDependency}(\mathit{task, children})$} \label{ln:cp-end}
\algline[2]{\textbf{else}} \label{ln:out-start}
\algline[3]{$\mathit{out}.\mathit{append}(\mathit{result})$} \label{ln:out-end}

\algline[1]{\textbf{if} $\mathit{result.filtered()=true}$ \textbf{or} $\mathit{stage} = \mathit{stages.lastStage}()$ \textbf{then}} \label{ln:release-start}
\algline[2]{$\mathit{memoryManager.releaseMemory}(\mathit{task})$} \label{ln:release-end}
\alglineplain{0}{} 

\algline{\textbf{procedure} \textsc{Rebalance}$(\mathit{stages})$}
\algline[1]{\textbf{for each} $\mathit{stage} \in \mathit{stages}$ \textbf{do}}
\algline[2]{\textbf{if} $\mathit{memoryManager.isSaturated(stage)}$ \textbf{and} $\mathit{stage} \neq \mathit{stages}[0]$ \textbf{then}} \label{ln:saturate-start}
\algline[3]{$\mathit{memoryManager.transferMemSaturation}(\mathit{lastStage})$}\label{ln:saturate-end}
\algline[1]{\textbf{for} $i \in [\mathit{stages.lastStage().id}, 1]$ \textbf{do}} \label{ln:starve-start}
\algline[2]{$\mathit{stage} = \mathit{stages}[i]$}
\algline[2]{$\mathit{prevStage} = \mathit{stages}[i-1]$}
\algline[2]{\textbf{if} $\mathit{memoryManager.isStarving(stage)}$ \textbf{then}}
\algline[3]{$\mathit{memoryManager.transferMemStarvation}(\mathit{prevStage})$} \label{ln:starve-end}
\alglineplain{0}{}

\end{pseudoalgorithm}
\Description{Pseudocode for the scheduler main loop, task admission, and adaptive stage-budget rebalancing.}
\end{figure}

Algorithm~\ref{alg:scheduler_pseudocode} illustrates the scheduling algorithm for a single pipeline.
Each stage maintains a \(\mathit{waiting}\) task queue per-stage (tasks awaiting to be launched) and a \(\mathit{running}\) task set (tasks with reserved budget currently executing).
The algorithm iterates until the query is completed, so there are no more waiting or running tasks.
In each iteration, the scheduler (1) admits waiting tasks whose stage has available memory given its budget, (2) processes any completed task, and (3) rebalances stage budgets.

The \textsc{Launch} procedure checks each stage's waiting queue and launches tasks only when the stage has enough remaining budget. 
The check is done by the memory manager, which keeps track of how much memory has been allocated to each task in each stage.

The \textsc{Complete} procedure deals with completed tasks.
First, as discussed in Section~\ref{subsec:mem-based-schedule}, each task is run with a maximum amount of tokens it can generate (details in Section~\ref{sec:robustness}). If the task generates more tokens, it is interrupted by the LLM serving engine.
In this case, the scheduler reruns the task with an increases token amount (lines~\ref{ln:retry-start}-\ref{ln:retry-end}).
If the task completed successfully and it did not include a predicate operator that filtered its input, the scheduler continues query execution.
If the stage is not the last stage in the pipeline and ends with a Cartesian product, the scheduler executes the Cartesian product to create dependent children tasks, taking the right-side table tuples as input, and enqueues them for the next stage (lines~\ref{ln:cp-start}-\ref{ln:cp-end}).
The Cartesian product can optionally execute a UDF to select a subset of tuples from the right-side table to join.
The memory manager keeps track of the dependencies between tasks and uses this information to release the memory for the parent task only when all the children tasks have completed.
If the stage is the last stage in the pipeline, the result is appended to the materialized output (lines~\ref{ln:out-start}-\ref{ln:out-end}).
Finally, if a task has been filtered or it belongs to the last stage, the memory of the task is released, possibly together with the budget of the parent tasks if all the other sibling tasks have completed too (lines~\ref{ln:release-start}-\ref{ln:release-end}).

The \textsc{Rebalance} procedure adaptively rebalances memory budgets among stages.
The budget for each stage is always larger than \(\mathit{minBudget}_s\), which is the amount of memory required to execute one task and keep its prefix cached.
This ensures that the query can make progress at each stage without requiring evictions.
Initially, all stages except the last are assigned a budget equal to \(\mathit{minBudget}_s\), while the last stage is assigned all the remaining memory. Rebalancing is triggered whenever a stage is detected to be starving or saturated based on the size of its waiting queue.

The memory manager keeps a low and high queue threshold for each stage $s$, initialized as:
\[
\mathit{high}_s = \alpha \cdot \frac{\mathit{budget}_s}{\mathit{minBudget}_s},
\qquad
\mathit{low}_s = \beta \cdot \frac{\mathit{budget}_s}{\mathit{minBudget}_s},
\]
where \(\mathit{budget}_s\) is the current memory budget assigned to $s$ and \(\alpha, \beta \in (0,1)\) are configurable static parameters with $\alpha > \beta$. 
The memory manager then classifies the stage $s$ as \emph{starving} if the size of its waiting queue is smaller than $\mathit{low}_s$ or \emph{saturated} if it is larger than $\mathit{high}_s$.

\begin{figure}[t]
    \centering
    \includegraphics[width=1 \linewidth]{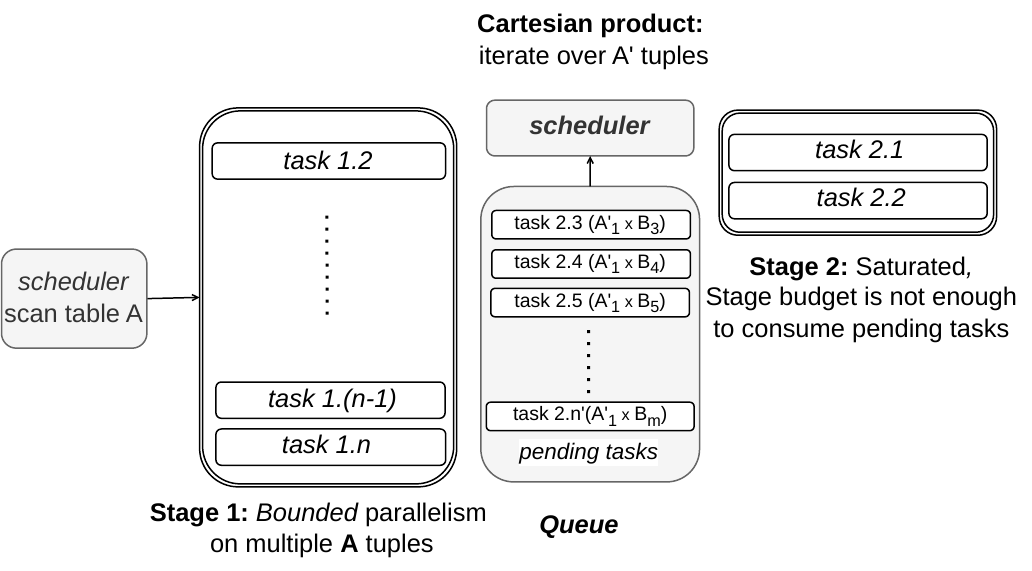}
    \caption{A parallel breadth-first policy can saturate the KV-cache budget.}
    \Description{Timeline showing a parallel breadth-first scheduling policy that admits too many early-stage tasks, fills memory with reserved prefix memory, and leaves insufficient capacity for downstream tasks.}
    \label{fig:saturation}
\end{figure}

\textsc{Rebalance} uses these states to transfer budget between stages.
If a stage is saturated (lines~\ref{ln:saturate-start}-\ref{ln:saturate-end}), this is an indication that its downstream stages cannot consume tuples at a sufficient rate, making tasks queue up.
Like in the example of Figure~\ref{fig:saturation}, the solution is to transfer memory from the saturated stage to the last stage.
This helps complete dependent tasks faster, evict cached prefixes, and use the budget that is now available to launch more waiting tasks.
The {\em transferMemSaturation} function transfers memory in blocks of size $\Delta$. 
If the saturated stage has at least $\Delta$ currently unused memory in his budget, $\Delta$ is transferred from the budget of that stage to the last stage.
Else, the algorithm iterates over all stages from the first one until it finds a stage with an unused block of size $\Delta$ and transfers from the budget of that stage.
The first stage is always saturated by design, so the algorithm does not apply this rule to it.
The memory manager transfers memory that has been just freed up by completed tasks.

If a stage is starving (lines~\ref{ln:starve-start}-\ref{ln:starve-end}),  this means that the upstream stages cannot produce enough tasks so they need a larger budget, similar to the situation depicted in Figure~\ref{fig:starvation}.
Therefore, the {\em transferMemStarvation} function transfers budget from the starving stage to the upstream stage.
It could happen that the upstream stage is also starving, in which case the algorithm iteratively transfers memory from downstream stages to upstream ones until it finds a non-starving stage.

\section{Memory Management}
\label{sec:robustness}
In \system, each task is assigned a maximum amount of tokens it can use, which is used to determine when it can be launched.
The previous discussion assumed, for simplicity, that memory estimation was perfect.
We now describe how the memory estimator component of \system predicts the expected memory usage of a task and how \system deals with incorrect estimates.
We also discuss dealing with deadlocks that can arise with explicit pinning.

\spara{Token bound estimation}
\label{sec:memory-estimation-reexecution}
Compared to traditional LLM serving workloads, relational workloads make it easier to predict the number of tokens generated by each LLM request, for multiple reasons.
Semantic operators often issue prefill-heavy requests, which generate few tokens or no tokens at all.
This is for example the case with filter operators, which have to output a binary value, and classifiers, which are used for aggregations and group-by operators.
More in general, semantic operators repeatedly execute the same instruction on many input tuples. The number of generated tuples per requests depends on the instruction and can be predicted through online monitoring, which is the approach used by \system.
We base our memory estimator on these insights.

For each stage, \system keeps a peak token usage bound for an entire task execution.
The user can decide whether to specify a static token bound for an operator, as is common for filter operators, or delegate the estimation of the bound to \system, which is useful for map-like operators that generate variable-length output. 
For the latter case, \system monitors completed requests and
records the ratio of generated output tokens to input (prompt) tokens. The
scheduler uses the calibrated 99th percentile of this
output-to-input token ratio as  the token bound for
future invocations of the same operator. This estimate keeps
the bound conservative enough to cover most requests while avoiding the
loss of parallelism caused by reserving memory for the worst case. 
If an LLM request terminates because it reaches its bound, \system interrupts the task, increases the bound for the task, and re-enqueues it. 

\spara{Explicit and virtual pinning}
A core principle of \system is to manage memory and schedule task launches to avoid evicting prefixes that are still needed.
\system offers two mechanisms to control evictions.
The first is \emph{explicit pinning} to prevent eviction of cached prefixes that will be reused by dependent tasks. 
This requires pinning support by the underlying LLM serving system.
The second is \emph{virtual pinning}, which is a best-effort alternative: the scheduler implicitly controls evictions by controlling which operators are launched and how much memory they use, assuming that the KV cache system uses an LRU eviction policy, and without explicit pinning.
The \system scheduling algorithm works with both variants.
\system offer both options because virtual pinning has some advantages over explicit pinning:
it does not require operators to implement a pinning logic or LLM serving systems to support pinning, and avoids memory deadlocks, a problem we now discuss.
In our evaluation, we found that virtual pinning achieves only slightly lower performance than explicit pinning.

\spara{Deadlock detection, avoidance, and recovery}
\enlargethispage{\baselineskip}
Explicit pinning can induce deadlocks when upstream tasks retain
memory needed by downstream tasks, but that pinned memory cannot be released
until the downstream tasks complete. In this setting, progress at one
stage can depend on memory held by another stage, creating a circular
wait across the pipeline. A similar situation can arise when the system retries interrupted tasks that exceeded their token bound. These tasks can further reduce the memory available to admit downstream work and can lead to deadlock.

The \system scheduling algorithm mitigates the risk of deadlocks by limiting how many concurrent tasks are launched.
Deadlocks, however, can still occur, so the system detects them by periodically querying the internal scheduler of the LLM serving engine and checking if there are waiting LLM requests but no running requests. Upon detection, \system unpins all memory and temporarily switches to virtual pinning, making it possible for the LLM engine to evict KV cache entries as needed.
In our experiments, we only observed deadlocks during stress-tests.

\section{Evaluation}
\label{sec:evaluation}

\noindent
We evaluate \system from two complementary perspectives. First, we
compare against existing semantic-operator systems~\cite{patel2025semantic,liu2025palimpzest} to measure end-to-end latency and
LLM invocation cost under comparable semantic tasks. Second, we study
\system's internal execution knobs, including pipelined versus blocking
execution, adaptive memory budgeting, virtual versus explicit pinning,
and output-token budget estimation, to understand which system
mechanisms drive the observed performance differences.

\subsection{Experimental Setup}

\spara{Baseline systems}
We compare our \system system with two baseline \semsyss that use request-centric LLM serving with operator-at-a-time execution. We configure all systems to use vLLM as the underlying LLM engine.
\begin{itemize}
	\item \emph{Lotus}~(v1.1.4)~\cite{patel2025semantic} proposes efficient implementations of those operators based on cascading: it first runs a cheap proxy implementation, which could use embedding similarity or a cheaper LLM model, and falls back to a more expensive reference LLM-based implementation if the proxy has low confidence. 
	Lotus runs user-defined query plans. 
	\item \emph{Palimpzest}~(v1.5.3)~\cite{liu2025palimpzest}, unlike Lotus, lets users specify a plan and then uses a query optimizer to improve it by reordering operators and selecting efficient implementations.
\end{itemize}

These systems focus on operator implementations and query optimization, which are complementary to \system's serving-layer optimizations.

\spara{System configuration}
All systems use the same operator implementations and manually-optimized query plan and oracle model (\nolinkurl{Llama-3.3-70B-Instruct}) for each workload, isolating the effect of the serving layer.
We detail the workload implementations in Section~\ref{sec:eval-workloads}.
For Lotus and Palimpzest, we use the default maximum batch size of 64.
For all systems, we set a static maximum token bound of 8 for predicate LLM calls, such as filter and join operators. This avoids cases where an LLM
non-deterministically produces many extra tokens for an otherwise short
operator output.
For Palimpzest, we disable \nolinkurl{reasoning_effort} to remove
few-shot reasoning prompts that would otherwise make Palimpzest prompts
longer, ensuring a fair comparison of serving performance.
For Lotus and \system, we evaluate also proxy operator implementations that use a smaller LLM than the default LLM. 
We configure them to use a separate vLLM instance running  the \nolinkurl{Llama-3.1-8B} model, which is small enough to fit in one GPU.
We do not use \system to optimize LLM serving for the small model.
We set the $\alpha$ and $\beta$ parameters used by \system for rebalancing (see Section~\ref{sec:scheduling-algo}) to $1$ and $0.5$ respectively.
We use virtual pinning by default.

\spara{Hardware and software configuration}
We run all experiments on a server with 4$\times$ NVIDIA RTX Pro 6000 GPUs (96 GB of HBM), and AMD EPYC 9575F Processor CPU with 8 cores and 300GB of RAM.
All systems we evaluate run on top of vLLM~v0.13.0rc4, with Automatic Prefix Caching (APC) enabled and the \nolinkurl{Llama-3.3-70B-Instruct} model distributed across all GPUs using tensor parallelism and the Triton attention backend.
By default, we assign 90\% of the total GPU memory to vLLM.
We configure vLLM to make query execution as deterministic as possible across multiple runs, following vLLM's reproducibility guidance~\cite{vllm_reproducibility_docs}.
We set \nolinkurl{VLLM_ENABLE_V1_MULTIPROCESSING=0}, which makes
offline scheduling deterministic. 
We use greedy decoding with temperature 0, top-$p$ 1.0, frequency penalty 0.5, and repetition penalty 1.3 to disable sampling and reduce repetitive generations.
This does not completely eliminate non-determinism~\cite{atil-etal-2025-non,yuan2026understanding}, so we also run each experiment three times and report average measurements.

\subsection{Workloads and Implementations}
\label{sec:eval-workloads}
We consider four workloads that exercise different operator combinations,
including both multi-operator multi-stage pipelines and single-stage
pipelines. 
We reimplement the same operator semantics on top of \system's API, including Lotus' proxies where applicable.
Our goal is to stress-test the performance of LLM serving and to show that it is possible to implement diverse operators and queries on top of \system's API.
A summary of the workloads and their datasets is shown in Table~\ref{tab:workload-datasets}.
We also report the average number of tokens per tuple and the average number of oracle LLM requests generated by the implementations in different systems, without considering proxy LLM calls for cascading implementations.
The numbers are slightly different across systems due to the inherent non-determinism of AI-based query execution.
The number in brackets for \system are the number of retried LLM requests due to incorrect memory estimations.
For the two workloads taken from the Lotus paper, FEVER and BioDEX, we follow the implementations described in~\cite{patel2025semantic}.

\begin{table}[t]
\centering
\small
\captionsetup{aboveskip=0pt}
\caption{Summary of Workloads.}
\label{tab:workload-datasets}
\renewcommand{\arraystretch}{1.08}
\setlength{\tabcolsep}{2.4pt}
\begin{tabular*}{\columnwidth}{@{\extracolsep{\fill}}p{0.14\columnwidth}p{0.12\columnwidth}>{\centering\arraybackslash}p{0.125\columnwidth}>{\centering\arraybackslash}p{0.10\columnwidth}>{\centering\arraybackslash}p{0.07\columnwidth}>{\centering\arraybackslash}p{0.07\columnwidth}>{\centering\arraybackslash}p{0.17\columnwidth}@{}}
\toprule
& & & & \multicolumn{3}{c@{}}{\# LLM calls} \\
\cmidrule(l){5-7}
Workload & Data & \# Tuples & \mbox{Avg. tok.} & Lotus & Palimp & \system \\
\midrule
FEVER &
  \begin{tabular}[c]{@{}l@{}}Claims\\Wikipedia\end{tabular} &
  \begin{tabular}[c]{@{}c@{}}1{,}000\\5,416,568\end{tabular} &
  \begin{tabular}[c]{@{}c@{}}11.8\\126.0\end{tabular} &
  1{,}183  & - & 1{,}243~(0) \\
\specialrule{0.1pt}{1pt}{1pt}
MEDEC &
  Patients &
  1{,}000 &
  397.1 &
  2,098 & 2,338 & 2{,}143~(21) \\
\specialrule{0.1pt}{1pt}{1pt}
BioDEX &
  \begin{tabular}[c]{@{}l@{}}Articles\\Reactions\end{tabular} &
  \begin{tabular}[c]{@{}c@{}}500\\11{,}271\end{tabular} &
  \begin{tabular}[c]{@{}c@{}}2{,}680.1\\6.3\end{tabular} &
  6,881 & - & 7{,}016~(0) \\
\specialrule{0.1pt}{1pt}{1pt}
  \begin{tabular}[c]{@{}c@{}}Contract\\NLI\end{tabular} &
  \begin{tabular}[c]{@{}l@{}}Contracts\\Hypotheses\end{tabular} &
  \begin{tabular}[c]{@{}c@{}}607\\17\end{tabular} &
  \begin{tabular}[c]{@{}c@{}}2{,}145.3\\36.8\end{tabular} &
  17,445 & 16,370 & 17{,}342~(167)\\
\bottomrule
\end{tabular*}
\renewcommand{\arraystretch}{1}
\end{table}

\begin{figure*}[!t]
	\centering
	\captionsetup{aboveskip=0pt,belowskip=-2pt}
	\includegraphics[width=.9\textwidth,trim=0 4pt 0 0,clip]{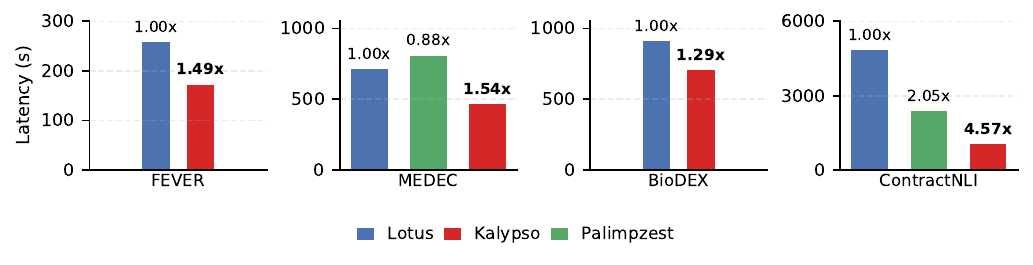}
	\caption{End-to-end latency and speedup across workloads.}
	\Description{Four bar plots showing end-to-end latency for FEVER, MEDEC, BioDEX, and ContractNLI. Each bar is annotated with speedup relative to Lotus.}
	\label{fig:end-to-end-latency}
\end{figure*}

\Needspace{16\baselineskip}
\spara{FEVER fact verification}
\label{sec:eval-fever}
\begin{workloadcode}
\codedf[claims]\semopmap(
    \codestr["Generate two concise Wikipedia search queries."]
)\indexopsearch(
    index\_table=\codedf[wikipedia_collection], top\_k=5
)\semopfilter(
    \codestr["Based on retrieved Wikipedia content, is the claim supported?"],
    cascade=True
)
\end{workloadcode}
FEVER is a fact verification workload, where each claim is checked against evidence taken from Wikipedia~\cite{thorne2018feverlargescaledatasetfact}. 
We sample 1{,}000 FEVER claims, as done in the Lotus paper~\cite{patel2025semantic}, and use that same query plan for all systems.
For each claim, the \texttt{sem\_map} operator generates
two concise search queries, which are issued to a ColBERT index search
built over a 5M-document Wikipedia corpus to retrieve candidate evidence
passages. 
This retrieval step is implemented as a tuple-independent
\texttt{index\_search} operator, which concatenates the retrieved evidence with the claim. 
The final \texttt{sem\_filter} operator verifies the claim based on the evidence.
It uses a small LLM as a proxy model, falling back to the oracle model if the proxy is not confident.
For Lotus, this implementation yields 183.3 oracle fallbacks over 1{,}000 claims. The \system implementation has 243.6 oracle fallbacks.
Palimpzest does not support merging the results of an index search into a single tuple so we don't include it in the evaluation of this workload.
Overall, this plan consists of a single-stage pipeline with three operators.

\vspace{1em}
\spara{MEDEC medical error detection and correction}
\begin{workloadcode}
\codedf[notes\_df]\semopfilter(
    \codestr["Answer true if any numbered sentence in the patient note "]
    \codestr["is medically inconsistent."]
)\semopmap(
    \codestr["Identify the numbered sentence that contains the medical error."]
)\semopmap(
    \codestr["Generate the corrected version of the erroneous sentence."]
) 
\end{workloadcode}
The MEDEC workload~\cite{abacha2025medecbenchmarkmedicalerror}
captures medical error detection and correction on clinical notes through three subtasks:
predicting whether the note contains a medical error,
identifying the sentence that contains the error when one is present,
and generating a corrected version of that sentence. We use a
1{,}000-patient sample from the MEDEC dataset.
We implement the query plan described in~\cite{abacha2025medecbenchmarkmedicalerror}: a \texttt{sem\_filter} operator performs error detection, the first
\texttt{sem\_map} operator identifies the erroneous sentence ID, and the
second one generates the corrected sentence. 
The Lotus implementation uses a cascading implementation with a smaller proxy LLM for the first filter.
We found that the behavior of the small LLM is highly non-deterministic for this workload, resulting in a high variance of oracle LLM calls across different runs.
Therefore, to ensure fairness, we use cascading only for Lotus and disable it for \system.
Although Palimpzest discusses cascades as a possible optimization, it does not yet implement the threshold-based fallback logic needed for these cascades, so we use an oracle-only implementation for Palimpzest too.
This plan has three operators and a single stage.

\vspace{1em}
\Needspace{16\baselineskip}
\spara{BioDEX biomedical reaction matching}
\begin{workloadcode}
\codedf[articles]\semopmap(
    \codestr["Extract adverse drug reaction labels described for the patient."]
)\semopjoin(
    right\_table=\codedf[reactions], icp=True
)\semopfilter(
    icp_predicate=\codestr["Does the article describe this adverse drug reaction?"],
    cascade\_low\_threshold=low, cascade\_high\_threshold=high    
)
\end{workloadcode}
\enlargethispage{2\baselineskip}
The BioDEX workload matches each biomedical article with the adverse drug reactions it describes~\cite{doosterlinck-etal-2023-biodex}.
The set of reactions is taken from a large database.
For all systems, we use the query plan and the operator implementations described in~\cite{patel2025semantic}.
Given a biomedical article, the \texttt{sem\_map} operator extracts
adverse drug reaction labels described in the article.
The \texttt{sem\_join} is implemented as an Index Cartesian Product (ICP) operator that retrieves matching candidate reactions from the reaction table using a vector index lookup with the \texttt{intfloat/e5-base-v2} embedding model.
The join then outputs all pairs of articles and retrieved candidate reactions, together with the vector distance between the two.
The following \texttt{sem\_filter} operator uses a proxy implementation that checks the vector distance of each pair.
Pairs with distances above a high threshold are rejected, those below
a low threshold are accepted, and uncertain candidates between the two
thresholds fall back to the oracle model.
We follow this implementation for both Lotus and \system.
Palimpzest does not support distance-based index search so we don't include it in the evaluation.
This plan consists of three operators and two stages, divided by the ICP operator.

\vspace{1em}
\spara{ContractNLI contract entailment}
\begin{workloadcode}
\codedf[contracts]\semopfilter(
    \codestr["Is this document a valid contract or agreement text "]
    \codestr["with enough clauses to evaluate confidentiality obligations?"]
)\semopjoin(
    right\_table=\codedf[hypotheses], \codestr["Given the contract and hypothesis, "]
    \codestr["does the contract entail the hypothesis?"]
)\semopmap(
    \codestr["Explain briefly why the contract entails the hypothesis, "]
    \codestr["citing the relevant obligation or clause."]
)
\end{workloadcode}
ContractNLI~\cite{koreeda2021contractnli} is a document-level
contract entailment workload: given a contract and a proposed legal statement (hypothesis), decide whether the contract supports (entails) that statement.
The \texttt{sem\_filter} operator removes documents that are not valid
contracts or do not contain enough substantive content to evaluate
confidentiality obligations. 
The \texttt{sem\_join} operator is a full Cartesian Product (CP) that pairs each
remaining contract with each hypothesis, followed by a semantic filter that checks whether the contract
entails that statement using an LLM. 
The final \texttt{sem\_map} operator generates a
concise explanation citing the relevant clause or obligation. 
For Lotus, we consider a cascading implementation with a smaller proxy LLM implementations for the first filter operators.
Like for MEDEC, we the small LLM shows a highly non-deterministic behavior that is hard to consistently reproduce, so we use an oracle-only implementation for \system.
We do the same for Palimpzest because it does not support cascading.
This plan has three operators and two stages, separated by the CP operator.

\begin{figure*}[!t]
    \centering
    \includegraphics[width=1\textwidth]{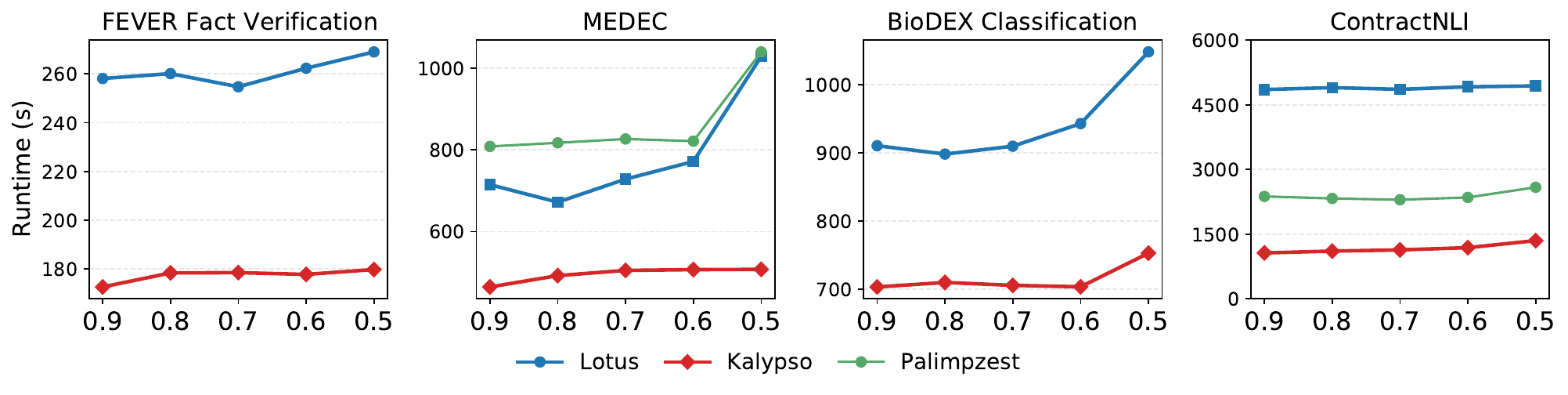}
    \caption{Effect of GPU memory utilization on end-to-end latency across workloads.}
    \Description{Line plots comparing Lotus, Palimpzest, and \system latency across FEVER, MEDEC, BioDEX, and ContractNLI workloads.}
    \label{fig:workload-gpu-utilization}
\end{figure*}

\subsection{End-to-End Query Completion Time}
\label{sec:eval-end-to-end}

Figure~\ref{fig:end-to-end-latency} summarizes the end-to-end query latency of
\system and the baseline semantic-query systems across the four
workloads. \system reduces latency for all workloads by executing semantic operators as a pipeline and by reusing KV-cache state across dependent operator calls, while the baselines execute operators one at a time and materialize intermediate results between operators.
LLM-call counts are similar across systems, as shown in Table~\ref{tab:workload-datasets}, so the speedup comes primarily from execution efficiency and depends on workload structure. 

Overall, the benefit depends on how much dependent LLM work can be
pipelined and how often downstream operators reuse the same tuple prefix.

For FEVER, \system reduces latency from 258.1s for Lotus to 172.6s, a
1.49$\times$ speedup. 
Palimpzest does not support this workload, as we described previously.
The workload has no Cartesian product, so the
entire memory budget is allocated to one stage. 
\system improves latency by pipelining the map, retrieval, and filter operators
and by preserving reusable KV-cache state across the dependent LLM
operators.
It controls concurrent task launches to ensure that each task can complete without having its KV cache state evicted.

MEDEC also has no Cartesian product and, compared to Fever, it has more downstream operators
that can reuse the prefix computed by the first operator.
The speedups of pipelining grow accordingly, reducing latency from
714.6s for Lotus and 808.2s for Palimpzest to 464.3s,
corresponding to 1.54$\times$ and 1.74$\times$ speedups. 
It is worth noting that \system only uses an oracle LLM, unlike Lotus which uses a cheaper but potentially unreliable proxy model for the filter.

BioDEX is a two-stage workload. 
We cannot run Palimpzest on this workload, as described previously.
\system reduces latency from 910.3s to 703.4s (1.29$\times$) by pipelining and reusing the prefixes for the articles across the ICP join and the final filter. Articles are long, averaging 2{,}680 tokens, which boosts speedups by increasing the benefit of prefix caching and reuse.
Gains are bounded by the use of a proxy implementation for the final filter operator, which skips LLM invocations in some cases based on the vector distances.
Although the ICP join can create many second-stage tasks, the size of the right-hand reaction tuples averages only 6.3 tokens.
Extra memory in this stage is therefore less effective.

ContractNLI is a two-stage workload combining long contract inputs, averaging
2{,}145 tokens, with 17 hypotheses per contract. The long contract
prefix can be reused by the join predicate and the final \texttt{sem\_map}. 
\system preserves this large reusable state across stage
boundaries, avoiding repeated prefill work. 
Against the lower-latency Palimpzest baseline, \system reduces latency from
2{,}373.4 to 1{,}062.4, a 2.23$\times$ speedup. Compared with
Lotus, the speedup is 4.57$\times$.
This is despite the fact that \system only uses an oracle LLM, whereas Lotus uses a proxy LLM for the filter operator.

\subsection{Execution Policy Ablations}
\label{sec:eval-ablation}

\begin{figure}[H]
    \centering
    \captionsetup{aboveskip=0pt,belowskip=-4pt}
    \includegraphics[width=.8\linewidth,trim=0 0 0 12,clip]{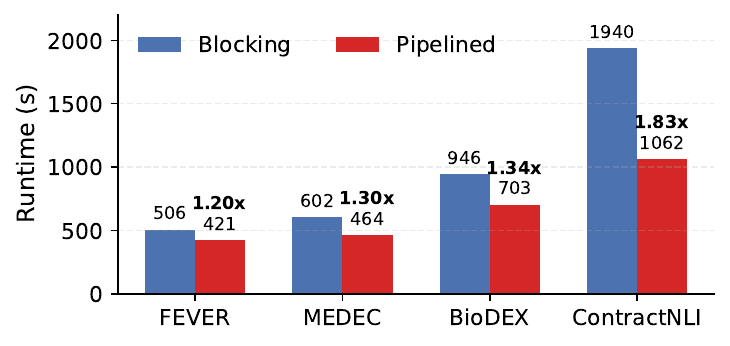}
    \caption{Latency for \system in blocking mode (operator-at-a-time, no cross-operator KV-cache reuse or pinning) vs.\ pipelined mode (default).}
    \Description{Grouped bar plot comparing blocking and pipelined execution latency across FEVER, MEDEC, BioDEX, and ContractNLI.}
    \label{fig:blocking-vs-pipelined}
\end{figure}

\spara{Blocking vs.\ pipelined execution}
\label{sec:eval-blocking-vs-pipelined}
To isolate the benefit of pipelined query execution over blocking, we modify \system to execute
operators in a blocking, operator-at-a-time manner. In this variant, each operator runs over the full input before
the next operator begins. As a result, downstream operators can no
longer inherit pinned KV-cache state from upstream execution and may
need to recompute the same tuple prefix instead of reusing it. 
Figure~\ref{fig:blocking-vs-pipelined} shows that pipelined execution is
faster on all four workloads, improving latency by 1.20--1.83$\times$ over
blocking execution. 
Across workloads, the gains reflect both pipeline overlap and
cross-operator prefix reuse. 
The speedup trends are consistent with the results shown previously.
These results therefore capture both
sources of improvement in \system's execution policy: pipelining increases overlap across operators, and
memory-aware scheduling preserves reusable KV-cache state long enough for downstream operators to benefit from it.
Speeups are larger for workloads where pipelined tuples are larger and pipelines consist of more operators.

\begin{figure}[t]
	\centering
	\captionsetup{aboveskip=2pt,belowskip=0pt}
	\includegraphics[width=1\linewidth]{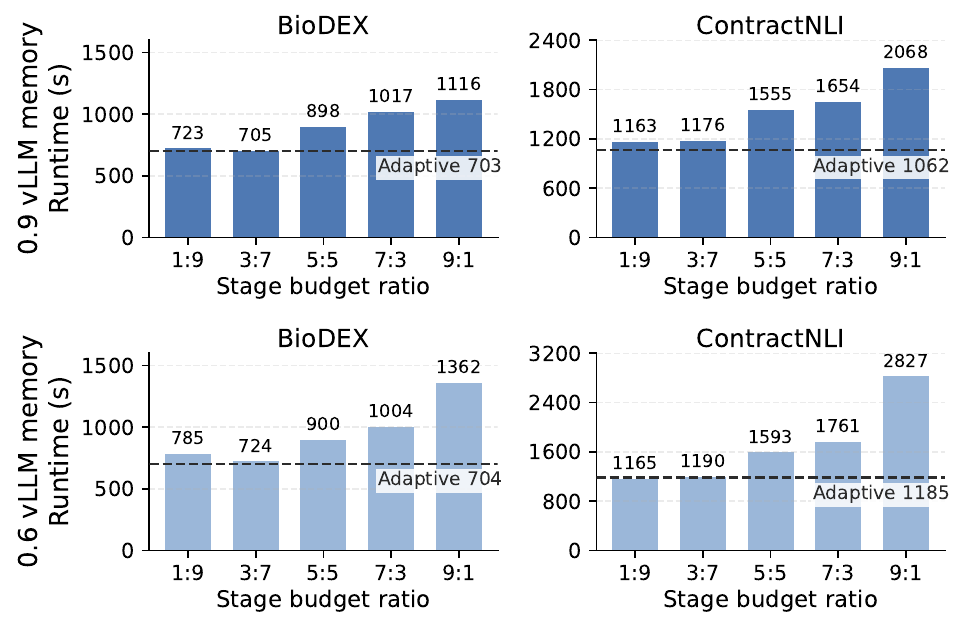}
\caption{End-to-end latency under fixed two-stage memory-allocation ratios and adaptive budgeting at 0.9 and 0.6 vLLM memory utilization.}
\Description{Two-by-two bar plots comparing fixed stage-budget ratios and adaptive budgeting on BioDEX and ContractNLI at 0.9 and 0.6 vLLM memory utilization.}
\label{fig:adaptive-budgeting}
\end{figure}

\spara{Impact of scheduling algorithms}
\label{sec:eval-adaptive-budgeting}
\system uses an adaptive algorithm to allocate memory budgets across stages.
We now show that a simpler static per-stage budget cannot consistently achieve similar performance across multiple workloads, as motivated in
Section~\ref{sec:scheduling-problem}. 
We use BioDEX and ContractNLI because they are multi-stage workloads
and compare adaptive budgeting against static two-stage memory-allocation ratios. 
In each static-ratio baseline, the scheduler assigns a constant fraction of the
available KV-cache budget to the first stage and the remainder to the
second stage throughout the query. We sweep the fixed split across the
ratios shown in Figure~\ref{fig:adaptive-budgeting}. 
All these baselines still rely on the same memory-aware scheduling approach introduced by \system: they organize operators in tasks and launch them only when memory is available, based on their estimated token bound.

Figure~\ref{fig:adaptive-budgeting} shows that no single static ratio is
best across workloads. 
The results also confirm that both starvation and saturation can impact system performance negatively.
ContractNLI performs best with a 10\%--90\%
allocation across the first and second stage, while BioDEX performs best with a 30\%--70\% allocation. The
best split therefore depends on workload-specific properties such as
tuple size, selectivity, join fanout, and the amount of downstream work
created by each stage. The impact of these factors is difficult to know before execution and can change as the query progresses. Adaptive budgeting avoids choosing this ratio ahead of time: it remains better than the best static allocation on both workloads by shifting memory toward the stage that is currently limiting pipeline progress. 

We repeated the experiment after reducing the memory available to vLLM to 60\% and found similar results.
The best runtime across static allocation ratios varies, and the adaptive strategy remains the best.

\spara{Virtual vs.\ explicit pinning}
The default for this paper's evaluation is to use virtual pinning, which preserves reusable prefixes through scheduling and admission control, rather than explicit pinning, which requires KV-cache pinning support from the LLM serving engine. We now compare the two variants.
Implementing explicit pinning required only modest changes to vLLM, totaling around 200 LoC across core vLLM components responsible for KV-cache management, block allocation, and request state. 
These changes prevent pinned KV-cache blocks from being evicted until they are explicitly unpinned by \system's scheduler.

\begin{figure}[!tbp]
\centering
\captionsetup{belowskip=0pt}
\includegraphics[width=.95\columnwidth]{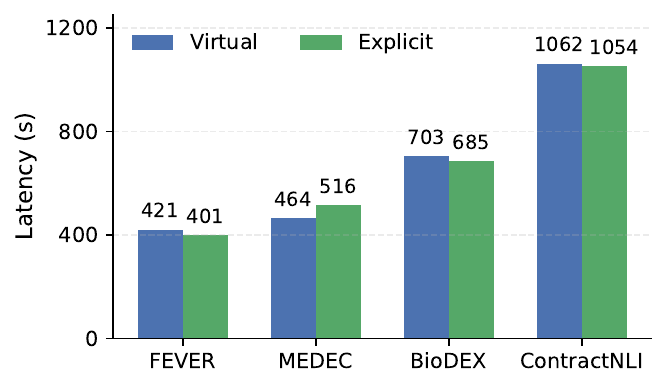}
\caption{Latency with virtual pinning and explicit pinning.}
\Description{Grouped bar plot comparing virtual pinning and explicit pinning. Explicit pinning is slightly faster on FEVER, BioDEX, and ContractNLI, but slower on MEDEC.}
\label{fig:explicit-pinning-ablation}
\end{figure}

Figure~\ref{fig:explicit-pinning-ablation} compares explicit pinning
against virtual pinning. The two modes have similar performance across
workloads: explicit pinning is slightly faster on FEVER, BioDEX, and
ContractNLI, while virtual pinning is faster on MEDEC. This result
suggests that virtual pinning is sufficient to capture the main
scheduling benefit without requiring explicit KV-cache pinning support
from the serving engine. Explicit pinning can still be used as an
optional backend feature when the serving engine provides it.

\spara{Token bound sensitivity}
\label{sec:eval-output-token-budget}
\system assigns a token bound to all LLM requests, which is the basis of memory-aware scheduling.
Its memory estimator avoids both overly conservative and overly optimistic bounds.
A worst-case budget reduces concurrency, while a very small budget can create retry overheads.

We evaluate output-token budget sensitivity on MEDEC by setting the GPU memory utilization limit to 0.9, which is the default, and 0.6, which induces additional memory pressure. 
As described in Section~\ref{sec:memory-estimation-reexecution}, the scheduler must
reserve KV-cache memory before knowing how many tokens a generative
operator will actually produce.

\begin{figure}[!t]
\centering
\captionsetup{aboveskip=4pt,belowskip=0pt}
\includegraphics[width=.95\columnwidth]{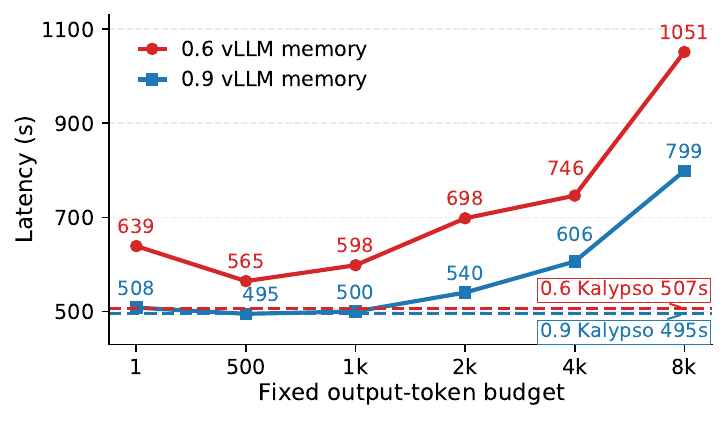}
\caption{Latency under fixed output-token budgets. The dashed line shows \system's default token budgeting.}
\Description{Line plot showing latency under fixed output-token budgets and \system's default token budgeting, which is lower than every fixed budget in the sweep.}
\label{fig:initial-output-token-budget}
\end{figure}

Figure~\ref{fig:initial-output-token-budget}
shows that conservative and overly large fixed budgets substantially increase latency because the scheduler admits fewer concurrent requests. 
At the other extreme, with a one-token generation budget, the scheduler allocates memory for each request's prefix, which includes data and instructions, and only one extra token.
The scheduler admits more concurrent requests, speeding up prefills, but it also needs to retry most requests.
The first execution of a request is fast, since it generates only one token.
For most requests, the subsequent retry execution finds the prefix still in the cache.
However, since re-executions use larger budgets, some prefixes might have been evicted.
\system's token budgeting achieves the best latency in the sweep without requiring the user to know the workload's output-token distribution in advance.

\spara{Resource sensitivity}
\label{sec:eval-resource-sensitivity}
We now stress the memory-management component of \system: as less KV-cache
memory is available, cached prefixes are more likely to be evicted unless
the scheduler preserves them for downstream operators.
We vary the GPU memory budget available to vLLM across five
settings: 0.9, 0.8, 0.7, 0.6, and 0.5, corresponding to approximately
190.6, 152.6, 114.6, 76.6, and 38.6~GB of KV-cache capacity. 
Figure~\ref{fig:end-to-end-latency} summarizes the high-memory setting,
while Figure~\ref{fig:workload-gpu-utilization} shows that \system remains fast
and robust as the KV-cache budget shrinks. Across all five memory
settings, \system is faster than the baselines on every workload. The
advantage is stable on FEVER and BioDEX, where reducing the budget does
not substantially change the relative ordering. Under stronger memory
pressure, \system also avoids the large slowdowns seen in the baselines:
on MEDEC, reducing the budget from 190.6~GB to 38.6~GB increases
\system's latency by only 9\% (464.3s to 507.3s), compared with 44\%
for Lotus and 29\% for Palimpzest. ContractNLI shows the same robustness
at larger scale: \system stays below 1348.3s across the sweep, while
Palimpzest remains above 2296.5s and Lotus remains above 4854.0s.

\section{Related Work}
\label{sec:related-work}
\spara{LLM Serving Systems}
Modern LLM serving systems optimize the execution of independent inference
requests. Orca introduces iteration-level scheduling and selective
batching~\cite{yu2022orca}; vLLM improves KV-cache memory efficiency
through PagedAttention~\cite{kwon2023efficient}; and Sarathi reduces
prefill--decode interference through chunked prefills
~\cite{agrawal2023sarathiefficientllminference}. DistServe disaggregates
prefill and decode across separately provisioned GPUs
~\cite{zhong2024distserve}, while Llumnix dynamically migrates requests
among model instances to improve load balance and isolation
~\cite{sun2024llumnix}. These systems schedule requests without visibility
into the semantic query plan. \system leverages these optimizations by runing on top LLM serving systems.
It uses operator dependencies to pipeline execution and coordinate KV-cache retention across related requests.

\spara{Prefix-Aware LLM Serving}
Several LLM serving systems exploit repeated prompt prefixes to avoid
recomputing KV-cache state. SGLang uses RadixAttention to organize cached
prefixes in a radix tree for structured and multi-turn language-model
programs~\cite{zheng2024sglang}, while vLLM supports automatic prefix
caching for requests with shared prompt prefixes
~\cite{kwon2023efficient}. Prompt-caching mechanisms similarly
reuse attention state for stable prompt components such as system
messages, templates, and long context documents~\cite{gim2024prompt}. These system support prefix reuse opportunistically among requests visible to them, as long as the relevant prefixes remain in they KV-cache. However, they do not control the higher-level workload that generates these requests, and therefore cannot actively restructure query execution to maximize cross-operator prefix reuse.
\system relies on these systems' prefix reuse capabilities to optimize pipelining.

\spara{Optimizing Semantic Operator Implementations}
Prior work reduces expensive LLM calls using model cascades and proxy
models~\cite{patel2025semantic,shankar2026task,Chung_2026,
zeighami2026bargain}. For semantic joins, feature decomposition extracts
relevant fields once per record and rewrites the join condition as a
logical expression over inexpensive feature comparisons, replacing
quadratic pairwise LLM evaluation while providing statistical precision
and recall guarantees~\cite{zeighami2025fdj}. Block-based joins instead place batches from both
inputs in each prompt, reducing the quadratic number of pairwise LLM
invocations, although batching can degrade output accuracy
~\cite{trummer2025semanticjoin}. Other systems organize document chunks
for prompt-cache discounts~\cite{shankar2026task} or fuse semantic
operators in streaming query
plans~\cite{chen2025continuouspromptsllmaugmentedpipeline}, but they do
not account for the serving engine memory pressure, parallelism, and scheduling.
\system complements them by optimizing the LLM serving layer.

\section{Conclusion}
\label{sec:conclusion}
We presented \system, a query-aware execution system for semantic
queries that bridges semantic query processing and LLM serving.
Existing semantic query systems execute operator invocations in a
query-agnostic manner, missing opportunities for cross-operator prefix
reuse and incurring unnecessary KV-cache eviction and recomputation
under memory pressure. By characterizing operators according to their
execution behavior, \system identifies pipelining opportunities,
executes query plans as stages and tasks, and manages KV-cache state
through memory-aware admission control.
These mechanisms enable \system to turn semantic query plans
into memory-aware execution pipelines that preserve reusable prefixes
across operators under bounded GPU memory.

\bibliographystyle{ACM-Reference-Format}
\bibliography{refs}

\end{document}